\newcommand{\CLASSINPUTtoptextmargin}{1.5in}
\newcommand{\CLASSINPUTbottomtextmargin}{1in}
\documentclass[conference,a4paper]{IEEEtran}
\ifCLASSINFOpdf
\else
\fi
\usepackage{amsmath,dsfont,bbm,epsfig,amssymb,amsfonts,amstext,verbatim}
\usepackage{amsopn,cite,subfigure}
\usepackage{multirow,multicol,lipsum,xfrac}
\usepackage{amsthm}
\usepackage{mathtools,amsthm}
\usepackage{perpage}
\usepackage{balance}
\usepackage{url}
\usepackage{amsfonts}
\usepackage{epsfig}
\usepackage[font={small}]{caption}
\usepackage{psfrag}	
\usepackage{etoolbox}
\usepackage{algorithmicx}
\usepackage[Algorithm,ruled]{algorithm}
\usepackage{algpseudocode}
\usepackage{pifont}
\usepackage[utf8]{inputenc}
\usepackage[T1]{fontenc}  
\usepackage[nolist]{acronym}
\usepackage{paralist}
\usepackage{enumitem}
\usepackage{bbm}
\usepackage[process=auto]{pstool}
\usepackage{tikz,pgfplots}
\usetikzlibrary{shapes,arrows}

\newcommand{\setR}{\mathbb{R}}

\newcommand{\setS}{\mathbbmss{S}}

\newcommand{\setC}{\mathbb{C}}

\newcommand{\rss}{\mathrm{RSS}}

\newcommand{\diag}[1]{\mathrm{diag} \left( #1 \right) }
\newcommand{\brc}[1]{\left( #1 \right) }
\newcommand{\dbc}[1]{\left[ #1 \right] }

\newcommand{\her}{\mathsf{H}}

\newcommand{\bvv}{\mathbf{v}}

\newcommand{\bh}{{\mathbf{h}}}
\newcommand{\bt}{{\mathbf{t}}}
\newcommand{\bp}{{\mathbf{p}}}

\newcommand{\bx}{{\boldsymbol{x}}}

\newcommand{\set}[1]{\left\lbrace#1\right\rbrace}

\newcommand{\bz}{{\boldsymbol{z}}}

\newcommand{\bs}{{\boldsymbol{s}}}

\newcommand{\bww}{\mathbf{w}}

\newcommand{\by}{{\boldsymbol{y}}}

\newcommand{\trp}{\mathsf{T}}

\newcommand{\mA}{\mathbf{A}}
\newcommand{\mW}{\mathbf{W}}

\newcommand{\mP}{\mathbf{P}}

\newcommand{\mI}{\mathbf{I}}

\newcommand{\mQ}{\mathbf{Q}}

\newcommand{\mV}{\mathbf{V}}

\newcommand{\mH}{\mathbf{H}}

\newcommand{\avg}{\mathrm{avg}}

\newcommand{\Ex}[2]{\mathbbmss{E}\left\lbrace #1 \right\rbrace_{#2}}

\newcommand{\sinr}{{\mathrm{SINR}}}

\newcommand{\norm}[1]{\lVert #1 \rVert}

\newcommand{\abs}[1]{\lvert #1 \rvert}
\newcommand{\tr}[1]{\mathrm{tr} \left\lbrace #1 \right\rbrace }

\newtheoremstyle{mystyle}
  {}
  {}
  {\it}
  {}
  {\bfseries}
  {:}
  { }
  {}

\theoremstyle{mystyle}

\algnewcommand\algorithmicLet{\textbf{Let}}
\algnewcommand\Let{\item[\algorithmicLet]}
\algnewcommand\algorithmicSet{\textbf{Set}}
\algnewcommand\Set{\item[\algorithmicSet]}

\algnewcommand\algorithmicInitiate{\textbf{Initiate}}
\algnewcommand\Initiate{\item[\algorithmicInitiate]}
\algnewcommand\algorithmicStart{\textbf{Begin}}
\algnewcommand\Begin{\item[\algorithmicStart]}
\algnewcommand\algorithmicEnd{\textbf{End}}
\algnewcommand\End{\item[\algorithmicEnd]}

\algnewcommand\algorithmicOutP{\textbf{Output:}}
\algnewcommand\Out{\item[\algorithmicOutP]}

\algnewcommand\algorithmicInP{\textbf{Input:}}
\algnewcommand\In{\item[\algorithmicInP]}

\algnewcommand\algorithmicFind{\textbf{Find}}
\algnewcommand\Find{\item[\algorithmicFind]}

\newcounter{bar}

\usepackage[a4paper,
            left=.92in,right=.9in,top=0.75in,
            bottom=1in,%
            ]{geometry}

\begin{document}
\title{Joint User Selection and Precoding in Multiuser MIMO Systems via Group LASSO}

\author{
\IEEEauthorblockN{
Saba Asaad\IEEEauthorrefmark{1},
Ali Bereyhi\IEEEauthorrefmark{1},
Ralf R. M\"uller\IEEEauthorrefmark{1}, and
Rafael F. Schaefer\IEEEauthorrefmark{2},\vspace*{1mm}
}
\IEEEauthorblockA{
\IEEEauthorrefmark{1}Institute for Digital Communications, Friedrich-Alexander Universit\"at Erlangen-N\"urnberg\\
\IEEEauthorrefmark{2}Information Theory and Applications Chair, Technische Universit\"at Berlin\\
\{saba.asaad, ali.bereyhi, ralf.r.mueller\}@fau.de, rafael.schaefer@tu-berlin.de
}
}
\IEEEoverridecommandlockouts

\maketitle

\begin{abstract}
Joint user selection and precoding in multiuser MIMO settings can be interpreted as group sparse recovery in linear models. In this problem, a~signal~with group sparsity is to be reconstructed from an underdetermined system of equations. This paper utilizes this equivalent interpretation and develops a computationally tractable algorithm based on the method of group LASSO. Compared to the state of the art, the proposed scheme shows performance enhancements in two different respects: higher achievable sum-rate and lower interference at the non-selected user terminals.\vspace*{2mm}
\end{abstract}

\IEEEpeerreviewmaketitle

\section{Introduction}
Performance gains are often achieved in multiuser massive \ac{mimo} systems with a large number of transmit antennas per user \cite{hoydis2013massive}. As a result, in dense settings in which the number of available users is comparable to the number of transmit antennas, user selection is required along with downlink beamforming \cite{dimic2005downlink,shen2005low,wang2008user,huang2013user}. 

The conventional approach for user selection and precoding is to divide them into two separate problems: First, a subset of users is selected; then, the information signals of~the~selected users are precoded via a classic precoding scheme \cite{huang2013user}.~Generally, the optimal approach for user selection deals with integer programming. Hence, this problem is often addressed via~sub-optimal greedy algorithms \cite{dimic2005downlink,shen2005low}. In this work, we deviate from the conventional approach and propose a scheme for joint user selection and downlink beamforming.

\subsection{User Selection and Precoding as Group Sparsity}
Joint user selection and beamforming~is interpreted as the problem of constructing a signal with \textit{group sparsity}.~To~clarify this point, assume a multiuser downlink scenario with $M$ transmit antennas and $K$ users in which we wish to select a subset of $L$ users. A linear precoder in this problem can be seen as a signal with $MK$ entries, such that each block of size $M$ represents an individual beamforming vector. By such a formulation, joint user selection and downlink beamforming with respect to some performance metric, e.g., the achievable sum-rate or \ac{mse}, reduces to the problem of finding a signal with group sparsity: {A signal of size $MK$~in which only $L$ blocks of size $M$ have non-zero entries}.

Following this equivalent interpretation, we employ the \ac{glse} framework for precoding, recently developed in \cite{bereyhi2017nonlinear,bereyhi2017asymptotics,bereyhi2018glse}, to formulate joint user selection and precoding as the problem of group sparse recovery~in~a linear model. A computationally tractable algorithm is~then developed based on group \ac{lasso} to address this problem. Our investigations show significant performance enhancements compared to the state of the art.

\subsection{Notations}
Throughout the paper, scalars, vectors, and matrices~are~represented by~non-bold, bold lower case, and bold upper case letters, respectively. The real axis is denoted by $\setR$ and the complex plane is shown by $\setC$. $\mH^{\her}$, $\mH^{*}$, and $\mH^{\trp}$ indicate the Hermitian,~complex~conjugate, and transpose of $\mH$, respectively. $\log\left(\cdot\right)$ is the binary logarithm. We denote the statistical expectation by $\Ex{\cdot}{}$. $\diag{\bt}$ represents the diagonal matrix constructed from the elements of vector $\bt$.

\section{Problem Formulation}
\label{sec:sys}
Consider a multiuser \ac{mimo} system with multiple \acp{bs} which are equipped with transmit~antenna~arrays of size $M$. The system is intended to serve $K$ single-antenna \acp{ut}. For mathematical tractability, we focus on a single \ac{bs} which aims to transmit information to a group of $L \leq K$ \acp{ut}. 

\subsection{System Model}
The system operates in the \ac{tdd} mode. Hence, the uplink and downlink channels are reciprocal. In each coherence time interval, the \acp{ut} transmit known training sequences. The \ac{bs} then utilizes these sequences~to~estimate the \ac{csi}.

Let $\bh_k\in\setC^{M}$ denote the vector of uplink channel coefficients between \ac{ut} $k$ and the \ac{bs}. The signal received by \ac{ut} $k$ is hence given by
\begin{align}
y_k = \bh_k^\trp \bx + z_k
\end{align}
where $z_k$ is additive complex Gaussian noise with zero mean and variance $\sigma_k^2$, i.e., $z_k \sim \mathcal{CN} \brc{0,\sigma_k^2}$, and $\bx$ is the downlink transmit signal constructed from the information symbols of the selected \acp{ut} and the \ac{csi}~via~linear~precoding. As a result, the transmit signal is written as
\begin{align}
\bx = \sum_{\ell \in \setS} \sqrt{p_\ell} s_\ell \bww_\ell.
\end{align}
where $\setS$, $s_\ell$, $p_\ell$ and $\bww_\ell$ are defined as follows:
\begin{enumerate}
\item $\setS\subseteq \set{1, \ldots,K}$ represents the subset of $L$ \acp{ut} selected by the \ac{bs} for downlink transmission.
\item $s_\ell$ is the information symbol of user $\ell$ which is assumed to be zero-mean and unit-variance.
\item $p_\ell$ denotes the power allocated to \ac{ut} $\ell\in\setS$.
\item $\bww_\ell$ is the beamforming vector of \ac{ut} $\ell$.
\end{enumerate}
The transmit power at the \ac{bs} is restricted. It is hence assumed that $\bx$ satisfies the power constraint $\Ex{\bx^\her \bx}{} \leq P$ for some non-negative real $P$.

\subsection{Performance Measure}
There are various metrics characterizing the performance of the downlink transmission in this system. One well-known metric is the \textit{weighted average throughput}  which is defined as
\begin{align}
R_\avg = \frac{1}{L} \sum_{\ell \in \setS} w_\ell R_\ell \label{eq:Throughput}
\end{align}
for some non-negative weights $\set{w_\ell}$~and~transmission rates 
\begin{align}
R_\ell = \log\brc{1+\sinr_\ell}. \label{eq:R_k}
\end{align}
In \eqref{eq:R_k}, $\sinr_\ell$ is defined as
\begin{align}
\sinr_\ell = \dfrac{\displaystyle p_\ell\abs{\bh_\ell^\trp \bww_\ell}^2}{\displaystyle \sigma_\ell^2 + \sum_{j=1,j\neq \ell}^K p_j\abs{\bh_\ell^\trp \bww_j}^2}.
\end{align}

From signal processing points of view, precoding can~be~interpreted as \textit{channel inversion}. In this problem, the ultimate~aim is to construct the transmit signal such that at a selected \ac{ut}~$\ell$, $\bh_\ell^\trp\bx = \beta s_\ell$, for some scaling factor $\beta$, and at \ac{ut} $k$ which has not been selected, we have $\bh_k^\trp\bx = 0$. The former guarantees channel inversion at the selected \acp{ut} which results in minimal post-processing load, and the latter restricts the precoder to have zero leakage at the non-selected \acp{ut}.

By this alternative viewpoint, a suitable performance measure  is the \textit{\ac{rss}} at the \acp{ut} defined~as
\begin{align}
\rss = \frac{1}{K} \sum_{k=1}^K \Ex{\abs{\bh_k^\trp\bx - \beta a_k s_k}^2}{},
\end{align}
where $a_k=1$ if \ac{ut} $k$ is selected and is zero otherwise.

\section{Optimal User Selection and Precoding}
Let $\bs = \dbc{s_1,\ldots,s_K}^\trp$ collect the~information~symbols of all \acp{ut}. By defining $p_k=0$ for those \acp{ut} which~are~not~selected, the transmit signal is compactly represented  as
\begin{align}
\bx = \mW \sqrt{\mP} \bs.
\end{align}
where $\mW$ and $\mP$ are defined as follows:
\begin{enumerate}
\item $\mW = \dbc{\bww_1,\ldots,\bww_K}$ is the beamforming matrix.
\item $\mP = \diag{\bp}$ with $\bp= \dbc{p_1,\ldots,p_K}^\trp$.
\end{enumerate}
The notation $\sqrt{\mP}$ moreover denotes a matrix whose entries are the square root of the entries of $\mP$. Similarly, the vector of receive signals $\by = \dbc{y_1,\ldots,y_K}^\trp$ reads
\begin{align}
\by = \mH^\trp \bx + \bz
\end{align}
where $\mH = \dbc{\bh_1,\ldots,\bh_K}$ and  $\bz= \dbc{z_1,\ldots,z_K}^\trp$.

\subsection{User Selection and Precoding with Minimum RSS}
We design the transmit signal by considering the \ac{rss} as the performance measure. In this respect, the optimal approach~for joint user selection and precoding is to find $\mW$ and $\bp$ such that the \ac{rss} is minimized and the signal constraints are satisfied. In the sequel, we formulate this approach in a standard form.

\subsubsection*{Objective Function}
Following the given representation, the \ac{rss} is written~as
\begin{align}
\rss = \frac{1}{K} \Ex{\norm{\mH^\trp \mW\sqrt{\mP} \bs - \beta \mA \bs}^2}{},
\end{align}
where $\mA=\diag{a_1, \ldots,a_K}$. In this formulation,~$\mA$ is ineffective and can be dropped. To show this,~note that for any non-selected \ac{ut} $k$, $\bx$ is independent of $s_k$ and hence
\begin{subequations}
\begin{align}
\hspace*{-2mm}\Ex{\abs{\bh_k^\trp\bx - \beta s_k}^2}{} \hspace*{-.7mm} &= \hspace*{-.7mm}\Ex{\abs{\bh_k^\trp\bx}^2}{} \hspace*{-.7mm}  + \hspace*{-.7mm} \beta^2 \Ex{\abs{ s_k}^2}{} \\
&= \hspace*{-.7mm}\Ex{\abs{\bh_k^\trp\bx}^2}{} \hspace*{-.7mm}+\hspace*{-.7mm} \beta^2.
\end{align}
\end{subequations}
Therefore, we can write
\begin{align}
\rss = \frac{1}{K} D\brc{\mW,\bp} - \brc{1-\frac{L}{K} } \beta^2,
\end{align}
where $D\brc{\mW,\bp}$ is defined as
\begin{subequations}
\begin{align}
D\brc{\mW,\bp} &\coloneqq \Ex{\norm{\mH^\trp \mW\sqrt{\mP} \bs - \beta \bs}^2}{}\\
&= \tr{ \mQ^\her \mQ }{}
\end{align}
\end{subequations}
with $\mQ = {\mH^\trp \mW\sqrt{\mP} - \beta \mI_K }$. We hence set the~objective~function to $D\brc{\mW,\bp}$.

\subsubsection*{Constraints}
There are two main constraints:
\begin{enumerate}
\item The number of selected \acp{ut} should be less than~$L$.
\item The average transmit power is constrained.
\end{enumerate}
Noting that the number of selected \acp{ut} in the system is given by the \textit{sparsity} of $\bp$, i.e., $\norm{\bp}_0$, the first constraint is written~as
\begin{align}
\norm{\bp}_0 \leq L.
\end{align}
For the second constraint, we note that
\begin{subequations}
\begin{align}
\Ex{\bx^\her \bx}{} &= \Ex{\bs^\her \sqrt{\mP} \mW^\her \mW \sqrt{\mP} \bs}{}\\
&\stackrel{\dagger}{=} \Ex{\tr{\sqrt{\mP} \mW^\her \mW \sqrt{\mP} \bs \bs^\her}}{}\\
&= \tr{\mW \mP \mW^\her}
\end{align}
\end{subequations}
where $\dagger$ follows the fact that $\Ex{{\bs\bs^\her}}{} = \mI_K$. As a result, the transmit power constraint reads 
\begin{align}
\tr{\mW \mP \mW^\her} \leq P.
\end{align}

\subsubsection*{Optimization Problem}
Considering the objective function and constraints, the jointly optimal approach for user selection and precoding is formulated as
\begin{alignat}{6}
&\min_{ \mW \in \setC^{M\times K} , \bp \in \setR_+^{K} }   &\qquad & D\brc{\mW,\bp} \label{eq:optProb_F}\\
&\mathrm{subject \ to} & &\mathrm{C_1:} \ \norm{\bp}_0 			\leq L,\nonumber\\
&                      & &\mathrm{C_2:} \ \tr{\mW \diag{\bp} \mW^\her}   \leq P.\nonumber
\end{alignat}
The optimization problem in its initial form is not tractable, since both the objective function and constraints~are~not~convex. We address this issue~by~converting \eqref{eq:optProb_F} into~a~\textit{group~selection} problem. We then develop an algorithm based on \textit{group \ac{lasso}} to estimate the solution.

\section{Precoding via Group LASSO}
The optimization problem in \eqref{eq:optProb_F} can be converted into a group selection problem. To show this, let $\mV \coloneqq \mW\sqrt{\mP}$ be the \textit{overall precoding matrix}. The objective function is rewritten in terms of $\mV$ as
\begin{align}
 D\brc{\mW,\bp} &= \tr{ \brc{\mH^\trp \mV- \beta \mI_K}^\her \brc{\mH^\trp \mV- \beta \mI_K} } \nonumber \\
&= \norm{ \mH^\trp \mV- \left. \beta \right. \mI_K }_{F}^2.
\end{align}
The power constraint is further given in terms of $\mV$~as
\begin{align}
\tr{\mV^\her \mV} = \norm{\mV}_F^2 \leq P.
\end{align}
To represent constraint $\rm C_1$ in terms of $\mV$, we note that only the column vectors in $\mV$ whose corresponding \ac{ut} is selected have non-zero entries. This equivalently means that
\begin{align}
\begin{cases}
\norm{\bvv_k} \neq 0 &\text{if \ac{ut} $k$ is selected}\\
\norm{\bvv_k} =    0 &\text{otherwise}
\end{cases},
\end{align}
where $\bvv_k = \sqrt{p_k} \bww_k$ denotes the $k$-th column vector of $\mV$.~As the result, one can write
\begin{align}
\norm{\mV}_{2,0} = \norm{\bp}_0,
\end{align}
where $\norm{\mV}_{p,q}$ denotes the $\ell_{p,q}$ norm of $\mV$ defined as
\begin{align}
\norm{\mV}_{p,q}  \coloneqq \dbc{ \sum_{k=1}^K \brc{\norm{\bvv_k}_p}^q }^{1/q}.
\end{align}

From the above derivations, we conclude that the optimal approach for joint user selection and precoding reduces to the following programming:
\begin{alignat}{6}
&\min_{ \mV \in \setC^{M\times K} }   &\qquad &\norm{ \mH^\trp \mV- \left. \beta \right. \mI_K }_{F}^2 \label{eq:optProb_F2}\\
&\mathrm{subject \ to} & &\mathrm{C_1:} \ \norm{\mV}_{2,0} 	\leq L,\nonumber\\
&                      & &\mathrm{C_2:} \ \norm{\mV}_F^2   \leq P.\nonumber
\end{alignat}

The optimization in \eqref{eq:optProb_F2} describes a group selection problem in which a matrix with \textit{group sparsity} is to be recovered, i.e., a matrix with a certain fraction of column or row vectors being zero. Such a problem raises in several applications,~e.g., distributed compressive sensing and machine leaning~\cite{sarvotham2005distributed,bereyhi2018theoretical,bach2008consistency}. Group selection in its primitive form is a \ac{np}-hard problem, since it reduces to an integer programming. To address this problem tractably, several suboptimal approaches have been developed in the literature which approximate the solution. Group \ac{lasso} is one of the most efficient approaches which relaxes the problem of group selection into a convex programming \cite{yuan2006model,deng2013group}. In the sequel, we use group \ac{lasso} to develop a computationally  tractable algorithm for joint user selection and precoding.

\subsection{A Tractable Algorithm via Group LASSO}
Group selection is an extension of~the basic \textit{sparse recovery} problem in which a sparse vector is to be recovered from an underdetermined system of equations \cite{donoho2006compressed,candes2008restricted}. Group \ac{lasso} extends Tibshirani's regularization approach \cite{tibshirani1996regression}~and convexifies the non-convex \textit{$\ell_0$-norm} with the \textit{$\ell_1$-norm}. This means that constraint $\rm C_1$ is relaxed as
\begin{align}
\mathrm{C_1:} \ \norm{\mV}_{2,1} 	\leq \eta L
\end{align}
for some $\eta$ which regularizes the relaxation. By doing so, the joint user selection and precoding reduces to
\begin{alignat}{6}
&\min_{ \mV \in \setC^{M\times K} }   &\qquad & \norm{ \mH^\trp \mV- \left. \beta \right. \mI_K }_{F}^2 \label{eq:optProb_F3}\\
&\mathrm{subject \ to} & &\mathrm{C_1:} \ \norm{\mV}_{2,1} 	\leq \eta L,\nonumber\\
&                      & &\mathrm{C_2:} \ \norm{\mV}_F^2   \leq P.\nonumber
\end{alignat}
This relaxed program represents a group \ac{lasso} algorithm which is convex and is posed as a generic linear programming.

\subsection{An Alternative Formulation via RLS}
The joint user selection and precoding scheme in \eqref{eq:optProb_F3} describes \textit{least~squares}~with side constraints, where the \ac{rss} $\norm{ \mH^\trp \mV- \left. \beta \right. \mI_K }_{F}^2$ is minimized subject to some constraints. Following~the~method of \textit{\ac{rls}}, this problem is converted into the following unconstrained optimization\footnote{Alternatively, one could use the method of Lagrange multipliers~to~conclude the similar unconstrained form.}
\begin{align}
\hspace*{-2mm}\min_{ \mV \in \setC^{M\times K} } \norm{ \mH^\trp \mV-  \beta  \mI_K }_{F}^2 +  \lambda  \norm{\mV}_F^2 +  \mu  \norm{\mV}_{2,1} \label{eq:RLS_final}
\end{align}
for some regularizers $\lambda$ a $\mu$. The key features of this algorithm are as follows:
\begin{itemize}
\item For given upper bounds on the group sparsity and transmit power of $\mV$, there exists a pair of regularizers $\lambda$ and $\mu$, such that the solution to \eqref{eq:RLS_final} satisfies the constraints. Hence, by tuning $\lambda$ and $\mu$ different constraints are fulfilled.
\item Due to its convexity, the problem is tractably solved via generic linear programming. Alternatively, an~iterative~algorithm based on \ac{amp} can be developed to find the solution with minimal~computational complexity; see \cite{rangan2011generalized} for more details on \ac{amp} and \cite{bereyhi2018precoding} for its applicatindons to precoding.
\end{itemize}

\begin{algorithm}[t]
\caption{Joint User Selection and Precoding}
\label{GroupLasso}
\begin{algorithmic}[0]
\In Channel matrix $\mH$, average transmit power $P$ and the number of selected users $L$.\vspace*{2mm}
\Set $\mV = \dbc{\bvv_1,\ldots,\bvv_K}$
\begin{align*}
\mV = \mathrm{GroupLASSO}\brc{\mH,P,L,\beta}
\end{align*}
\Let subset $\setS \subseteq \set{1,\ldots,K}$ contain indices of the column vectors in $\mV$ which have the $L$ largest $\ell_2$-norms, i.e., $\abs{\setS}=L$ and
\begin{align*}
\norm{\bvv_\ell}^2 \geq \norm{\bvv_j}^2
\end{align*}
for any $\ell\in\setS$ and $j\in \set{1,\ldots,K}-{\setS}$.
\Set $\bvv_j = 0$ for $j\in \set{1,\ldots,K}-{\setS}$, and update $\mV$ as
\begin{align*}
\mV \leftarrow \frac{\sqrt{P}}{\norm{\mV}_F} \left. \mV \right.
\end{align*}
\Set $p_k = \norm{\bvv_k}^2$ and $\bww_k = \dfrac{\bvv_k}{ \norm{\bvv_k}}$ for $k\in\set{1,\ldots,K}$.\vspace*{2mm}
\Out Beamforming matrix $\mW = \dbc{ \bww_1,\ldots,\bww_K }$ and power allocation matrix $\mP = \diag{p_1,\ldots,p_K}$.
\end{algorithmic}
\end{algorithm}

Using either the algorithm in \eqref{eq:optProb_F3} or the one in \eqref{eq:RLS_final}, a matrix $\mV$ is tractably found which approximates the optimal~solution to \eqref{eq:optProb_F2}. The beamforming and power allocation matrices are then given by decomposing this matrix as $\mV = \mW \sqrt{\mP}$ for a diagonal $\mP$. In the sequel, we investigate the performance of the proposed approach through some numerical simulations.

\section{Performance Investigation}
We study the performance of the proposed approach by simulating some sample scenarios. To jointly precode and select user via group \ac{lasso}, Algorithm~\ref{GroupLasso} is used. In this algorithm, 
\begin{align}
\mV = \mathrm{GroupLASSO}\brc{\mH,P,L,\beta}
\end{align}
denotes the solution to the minimization in \eqref{eq:optProb_F3} with $\eta=1$. The algorithm finds first the solution $\mV$ to \eqref{eq:optProb_F3}, and selects $L$ \acp{ut} with strongest precoding vectors while setting the other column vectors zero. It then scales the precoding vectors of the selected users, such that the downlink transmit signal remains $P$.

As a benchmark, we evaluate the performance of \ac{mrt} beamforming with random~user~selection, and compare it with the performance of Algorthm~\ref{GroupLasso}. In this approach, $L$ \acp{ut} are selected at random. The precoding vector of selected user $k$ is then set to 
\begin{align}
\bvv_k = \sqrt{\frac{P}{L}} \left. \frac{ \bh_k^* }{ \norm{\bh_k} } \right. .
\end{align}

Throughout the simulations the standard Rayleigh model~is considered for the fading channel. This means that the entries of $\mH$ are generated independently and identically~with~complex zero-mean and unit-variance Gaussian distribution,~i.e., 
\begin{align}
h_{mk} \sim \mathcal{CN}\brc{0,1}
\end{align}
for $m\in\set{1,\ldots,M}$ and $k\in\set{1,\ldots,K}$.

\subsection{Performance Metrics}
To quantify the performance, the following metrics are considered:
\begin{enumerate}
\item The weighted average throughput $R_\avg$ defined in \eqref{eq:Throughput} for uniform wights, i.e., $w_1, \ldots,w_K = 1$. This metric~determines the average achievable rate per selected \ac{ut} which is widely used in this literature.
\item The \textit{power leakage} to the non-selected \acp{ut} which is given by
\begin{subequations}
\begin{align}
Q_{\rm Leak} &\coloneqq \Ex{ \sum_{k=1, k\notin \setS}^K \abs{\bh_k^\trp\bx}^2 }{}\\
&= \sum_{k=1, k\notin \setS}^K \left. \sum_{\ell\in\setS} \right. \abs{\bh_k^\trp\bvv_\ell}^2.
\end{align}
\end{subequations}
This metric calculates the total amount of interference at the non-selected \acp{ut} from the downlink transmission to the selected \acp{ut}.
\end{enumerate}

\subsection{Scenario A: Fixed Loads}
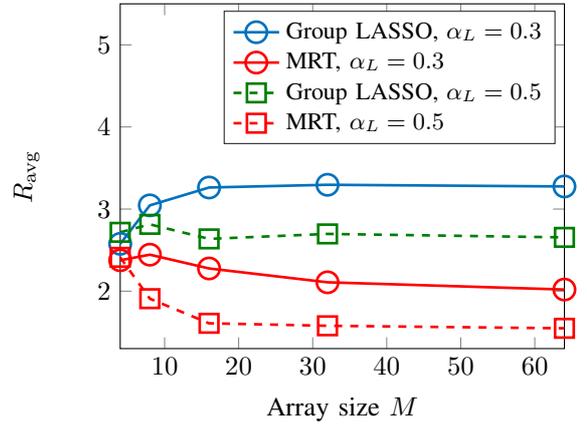
\begin{figure}[t]
\centering
%
%
\definecolor{mycolor1}{rgb}{0.00000,0.44706,0.74118}%
\definecolor{mycolor2}{rgb}{0.00000,0.49804,0.00000}%
\begin{tikzpicture}

\begin{axis}[%
width=2.3in,
height=1.8in,
at={(1.962in,0.942in)},
scale only axis,
xmin=4,
xmax=64,
xtick={10,20,30,40,50,60},
xticklabels={{$10$},{$20$},{$30$},{$40$},{$50$},{$60$}},
xlabel={Array size $M$},
ymin=1.3,
ymax=5.5,
ytick={1,2,3,4,5},
yticklabels={{$1$},{$2$},{$3$},{$4$},{$5$}},
ylabel={$R_\avg$},
axis background/.style={fill=white},
legend style={legend cell align=left, align=left, draw=white!15!black}
]
\addplot [color=mycolor1, line width=1.0pt, mark size=4.0pt, mark=o, mark options={solid, mycolor1}]
  table[row sep=crcr]{%
4	2.57617696280114\\
8	3.04497054246665\\
16	3.26256045146908\\
32	3.29561856702095\\
64	3.27520747804171\\
};
\addlegendentry{\small{Group LASSO, $\alpha_{ L} = 0.3$}}

\addplot [color=red, line width=1.0pt, mark size=4.0pt, mark=o, mark options={solid, red}]
  table[row sep=crcr]{%
4	2.37583185748851\\
8	2.44419074983416\\
16	2.27631182271319\\
32	2.10774437799402\\
64	2.01949019597847\\
};
\addlegendentry{\small{MRT, $\alpha_{ L} = 0.3$}}

\addplot [color=mycolor2, dashed, line width=1.0pt, mark size=3.5pt, mark=square, mark options={solid, mycolor2}]
  table[row sep=crcr]{%
4	2.71663841668229\\
8	2.81381931801315\\
16	2.63667882412798\\
32	2.69724148020715\\
64	2.65503917682481\\
};
\addlegendentry{\small{Group LASSO, $\alpha_{ L} = 0.5$}}

\addplot [color=red, dashed, line width=1.0pt, mark size=3.5pt, mark=square, mark options={solid, red}]
  table[row sep=crcr]{%
4	2.4116952938163\\
8	1.90843224849648\\
16	1.60837190302342\\
32	1.5780201728742\\
64	1.54749623661985\\
};
\addlegendentry{\small{MRT, $\alpha_{ L} = 0.5$}}

\end{axis}
\end{tikzpicture}%
\caption{Average throughput vs. the array size $M$. Here, $P=1$ and $\sigma_k^2 = 0.1$ for all the \acp{ut}. The user load is set to $\alpha_K=1$, and the scaling factor reads $\beta=1$.}
\label{fig:1}
\end{figure}

\begin{figure}[t]
\centering
%
%
\definecolor{mycolor1}{rgb}{0.00000,0.44706,0.74118}%
\definecolor{mycolor2}{rgb}{0.00000,0.49804,0.00000}%
\begin{tikzpicture}

\begin{axis}[%
width=2.3in,
height=1.8in,
at={(1.962in,0.942in)},
scale only axis,
xmin=4,
xmax=64,
xtick={10,20,30,40,50,60},
xticklabels={{$10$},{$20$},{$30$},{$40$},{$50$},{$60$}},
xlabel={Array size $M$},
ymode=log,
ymin=0.02,
ymax=25,
ytick={0.01,0.1,1,10},
yticklabels={{$10^{-2}$},{$10^{-1}$},{$10^{0}$},{$10^{1}$}},
ylabel={$Q_{\rm Leak}$},
yminorticks=true,
axis background/.style={fill=white},
legend style={legend cell align=left, align=left, draw=white!15!black}
]
\addplot [color=mycolor1, line width=1.0pt, mark size=4.0pt, mark=o, mark options={solid, mycolor1}]
  table[row sep=crcr]{%
4	0.0736135609209633\\
8	0.0690694655760798\\
16	0.0682345101559093\\
32	0.0686170326132175\\
64	0.0765943535021438\\
};
\addlegendentry{\small{Group LASSO, $\alpha_{L} = 0.3$}}

\addplot [color=red, line width=1.0pt, mark size=4.0pt, mark=o, mark options={solid, red}]
  table[row sep=crcr]{%
4	1.09408690027045\\
8	1.02888070329994\\
16	0.964798611328535\\
32	0.987572209943837\\
64	1.01263660906415\\
};
\addlegendentry{\small{MRT, $\alpha_{L} = 0.3$}}

\addplot [color=mycolor2, dashed, line width=1.0pt, mark size=3.5pt, mark=square, mark options={solid, mycolor2}]
  table[row sep=crcr]{%
4	0.0698504512593267\\
8	0.0678642742885117\\
16	0.0740178700392499\\
32	0.0776400513633597\\
64	0.0788401984164482\\
};
\addlegendentry{\small{Group LASSO, $\alpha_{L} = 0.5$}}

\addplot [color=red, dashed, line width=1.0pt, mark size=3.5pt, mark=square, mark options={solid, red}]
  table[row sep=crcr]{%
4	1.05798102840615\\
8	0.960548977196089\\
16	1.05892911027383\\
32	1.01128987503586\\
64	1.01407956973412\\
};
\addlegendentry{\small{MRT, $\alpha_{L} = 0.5$}}

\end{axis}
\end{tikzpicture}%
\caption{Power leakage vs. the number of transmit antennas $M$. Here, $P=1$ and $\sigma_k^2 = 0.1$ for all the \acp{ut}. The user load is set to $\alpha_K=1$, and the scaling factor reads $\beta=1$.}
\label{fig:2}
\end{figure}
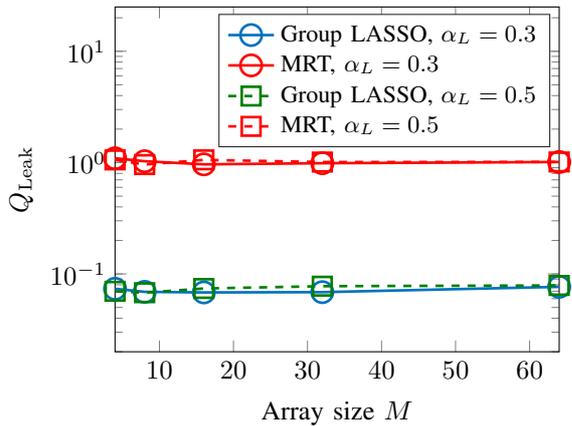

We first consider a scenario in which the total number of \acp{ut}, as well as the number of selected ones, is a fixed fraction of the transmit array size $M$. More precisely, a downlink transmission scenario is considered in which $K=\lceil \alpha_{ K} M \rceil$ number of users are available and we intend to select $L = \lceil \alpha_{ L} M \rceil$ \acp{ut}. Here, $\alpha_{ K}$ and $\alpha_{ L}$ are fixed numbers. For this scenario, both the performance metrics are sketched for fixed transmit power $P$ and noise variance in Fig.~\ref{fig:1} and Fig.~\ref{fig:2} in terms~of~the downlink transmit array size $M$.

Fig.~\ref{fig:1} shows the weighted average throughput\footnote{Remember that the average throughput in this case is defined as the sum-rate divided by the number of selected users.} against $M$. Here, $P=1$ and the noise variances are set to $\sigma_k = 0.1$ for $k\in\set{1,\ldots,K}$. Moreover, the scaling factor reads $\beta=1$. The results are sketched for $\alpha_K = 1$ and two different values of $\alpha_L$; namely, $\alpha_L\in\set{0.3,0.5}$. As the figure depicts, the proposed approach considerably outperforms the conventional \ac{mrt} technique. Such an enhancement comes from the joint selection and precoding approach. The convergence of $R_\avg$ to a constant in both the techniques follows hardening of the channel in large dimensions for fixed loads \cite{asaad2018massive,hoydis2013massive}.

The power leakage for this scenario is plotted in Fig.~\ref{fig:2}~versus $M$. Here, the parameters are set exactly to~the~ones~considered in Fig.~\ref{fig:1}. The figure~demonstrates the following~two observations:
\begin{enumerate}
\item The proposed algorithm imposes significantly less interference to the non-selected \acp{ut}. This observation comes from the fact that the objective function in \eqref{eq:optProb_F3} contains the power leakage as a penalty term.
\item The power leakage in both techniques converges to a constant value. Such a behavior is naturally following the fact that the loads $\alpha_K$ and $\alpha_L$ are kept fixed.
\end{enumerate}

\subsection{Scenario B: Fixed Number of UTs}
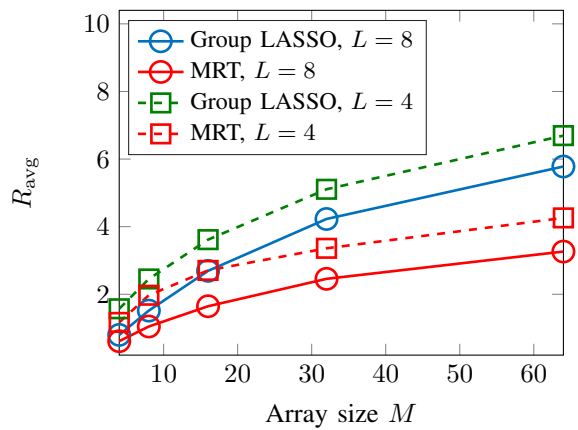
\begin{figure}[t]
\centering
%
%
\definecolor{mycolor1}{rgb}{0.00000,0.44706,0.74118}%
\definecolor{mycolor2}{rgb}{0.00000,0.49804,0.00000}%
\begin{tikzpicture}

\begin{axis}[%
width=2.3in,
height=1.8in,
at={(1.962in,0.942in)},
scale only axis,
xmin=4,
xmax=64,
xtick={10,20,30,40,50,60},
xticklabels={{$10$},{$20$},{$30$},{$40$},{$50$},{$60$}},
xlabel={Array size $M$},
ymin=0.2,
ymax=10.4,
ytick={2,4,6,8,10},
yticklabels={{$2$},{$4$},{$6$},{$8$},{$10$}},
ylabel={$R_\avg$},
axis background/.style={fill=white},
legend style={at={(0.02,0.58)}, anchor=south west, legend cell align=left, align=left, draw=white!15!black}
]
\addplot [color=mycolor1, line width=1.0pt, mark size=4.0pt, mark=o, mark options={solid, mycolor1}]
  table[row sep=crcr]{%
4	0.798185029052656\\
8	1.51713842624023\\
16	2.69264104312833\\
32	4.22733194171147\\
64	5.77993254062501\\
};
\addlegendentry{\small{Group LASSO, $L=8$}}

\addplot [color=red, line width=1.0pt, mark size=4.0pt, mark=o, mark options={solid, red}]
  table[row sep=crcr]{%
4	0.61186740057947\\
8	1.04730310372308\\
16	1.64573180877455\\
32	2.4566201196191\\
64	3.26391965855303\\
};
\addlegendentry{\small{MRT, $L=8$}}

\addplot [color=mycolor2, dashed, line width=1.0pt, mark size=3.5pt, mark=square, mark options={solid, mycolor2}]
  table[row sep=crcr]{%
4	1.57062447426457\\
8	2.45290032959929\\
16	3.6163038201864\\
32	5.10630189726423\\
64	6.6937760626111\\
};
\addlegendentry{\small{Group LASSO, $L=4$}}

\addplot [color=red, dashed, line width=1.0pt, mark size=3.5pt, mark=square, mark options={solid, red}]
  table[row sep=crcr]{%
4	1.16750228876215\\
8	1.97190076888829\\
16	2.70639862495651\\
32	3.36020932579522\\
64	4.25937991907839\\
};
\addlegendentry{\small{MRT, $L=4$}}

\end{axis}
\end{tikzpicture}%
\caption{Average throughput vs. the array size $M$. Here, $P=1$ and $\sigma_k^2 = 0.1$ for all the \acp{ut}. The number of \acp{ut} is set to $K=16$, and the scaling factor reads $\beta=1$.}
\label{fig:3}
\end{figure}

\begin{figure}[t]
\centering
%
%
\definecolor{mycolor1}{rgb}{0.00000,0.44706,0.74118}%
\definecolor{mycolor2}{rgb}{0.00000,0.49804,0.00000}%
\begin{tikzpicture}

\begin{axis}[%
width=2.3in,
height=1.8in,
at={(1.962in,0.942in)},
scale only axis,
xmin=4,
xmax=64,
xtick={10,20,30,40,50,60},
xticklabels={{$10$},{$20$},{$30$},{$40$},{$50$},{$60$}},
xlabel={Array size $M$},
ymode=log,
ymin=5e-14,
ymax=100,
ytick={1e-12,1e-08,0.0001,1},
yticklabels={{$10^{-12}$},{$10^{-8}$},{$10^{-4}$},{$10^{0}$}},
ylabel={$Q_{\rm Leak}$},
yminorticks=true,
axis background/.style={fill=white},
legend style={at={(0.02,0.02)}, anchor=south west, legend cell align=left, align=left, draw=white!15!black}
]
\addplot [color=mycolor1, line width=1.0pt, mark size=4.0pt, mark=o, mark options={solid, mycolor1}]
  table[row sep=crcr]{%
4	0.402598977130243\\
8	0.210659290184019\\
16	0.0738733278406697\\
32	5.36828916835021e-05\\
64	6.18354481699436e-13\\
};
\addlegendentry{\small{Group LASSO, $L=8$}}

\addplot [color=red, line width=1.0pt, mark size=4.0pt, mark=o, mark options={solid, red}]
  table[row sep=crcr]{%
4	1.01717768928374\\
8	0.936698079087211\\
16	0.967776444902207\\
32	1.06681735988472\\
64	1.04630551480431\\
};
\addlegendentry{\small{MRT, $L=8$}}

\addplot [color=mycolor2, dashed, line width=1.0pt, mark size=3.5pt, mark=square, mark options={solid, mycolor2}]
  table[row sep=crcr]{%
4	0.417123246763884\\
8	0.17336882191629\\
16	0.0655737069412962\\
32	3.01328717617722e-05\\
64	1.96775480527187e-12\\
};
\addlegendentry{\small{Group LASSO, $L=4$}}

\addplot [color=red, dashed, line width=1.0pt, mark size=3.5pt, mark=square, mark options={solid, red}]
  table[row sep=crcr]{%
4	1.0036955927201\\
8	0.992007031669601\\
16	0.973873310558162\\
32	0.97372682588975\\
64	0.942083968799431\\
};
\addlegendentry{\small{MRT, $L=4$}}

\end{axis}
\end{tikzpicture}%
\caption{Power leakage vs. the number of transmit antennas $M$. Here, $P=1$ and $\sigma_k^2 = 0.1$ for all the \acp{ut}. The number of \acp{ut} is set to $K=16$, and the scaling factor reads $\beta=1$.}
\label{fig:4}
\end{figure}
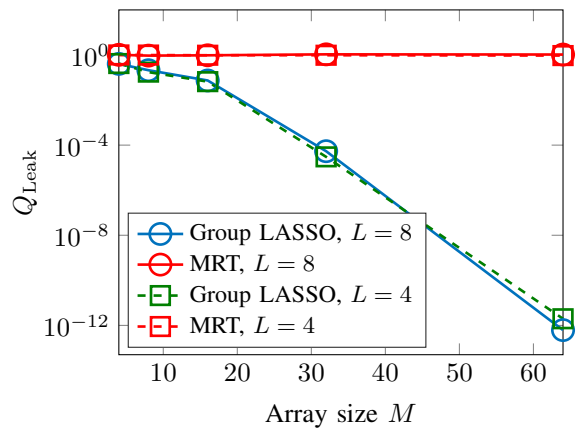

As another scenario, we consider a case in which the total number of \acp{ut}, as well as the number of selected ones, does not grow with $M$. For this case, we study a settings in which a downlink array of size $M$ is employed to service $L$ users out of $K=16$ available \acp{ut}. Similar to Scenario A, we set $P$ and noise variances to fixed numbers and sketch the average throughout, as well as the power leakage, against the transmit array size $M$ in Fig.~\ref{fig:3} and Fig.~\ref{fig:4}.

In Fig.~\ref{fig:3}, the average throughput $R_\avg$ is sketched against $M$ assuming $\beta=1$, $P=1$ and $\sigma_k=0.1$ for $k\in\set{1,\ldots,K}$. The results are given for $L\in\set{4,8}$. Similar to Scenario~A, the figure depicts performance enhancement achieved by using the proposed algorithm based on the group \ac{lasso}. In contrast to Scenario A, the throughput in this case grows logarithmically with $M$. Such a behavior follows the fact that in this case, the number of \acp{ut} is constant and does not grow with~$M$.

Fig.~\ref{fig:4} shows the variation of the power leakage against $M$. As the figure demonstrate, in the proposed algorithm, $Q_{\rm Leak}$ vanishes significantly fast as $M$ grows, such that at $M=64$ it imposes almost no interference to the non-selected \acp{ut}. Such a behavior follows the fact that in the joint approach based on the group \ac{lasso}, the beamforming vectors are constructed, such that the power leakage is suppressed at non-selected \acp{ut}. For a fixed number of \acp{ut}, the suppression is performed more accurately by narrow beamforming towards the selected users, as the array size grows large \cite{bereyhi2018robustness}.

\section{Conclusions}
A joint user selection and precoding scheme has been proposed for multiuser \ac{mimo} systems based on group \ac{lasso}. The scheme depicts performance enhancement in two different aspects: 
\begin{inparaenum}
\item The throughput of the system, defined as the sum-rate divided by the number of active users, shows some gains.
\item The interference imposed by downlink transmission at the non-selected \acp{ut} is significantly reduced. For instance,~when $L=8$ \acp{ut} are selected out of $K=16$ users,~there~is~almost zero interference, when the \ac{bs} is equipped with $M=64$ antennas.
\end{inparaenum}
These observations indicate that the proposed scheme is a good candidate for  massive \ac{mimo} settings.

The current work can be pursued in various directions. For example, considering the \ac{rls}-based derivation in \eqref{eq:RLS_final},~an~iterative algorithm can be developed via \ac{amp} implementing~the proposed scheme with low computational complexity. Another direction is to extend the current~framework~to wiretap settings following the approach in \cite{asaad2019asilomar}. The work in these directions is currently ongoing.

\bibliography{ref}
\bibliographystyle{IEEEtran}

\begin{acronym}
\acro{mimo}[MIMO]{multiple-input multiple-output}
\acro{mimome}[MIMOME]{multiple-input multiple-output multiple-eavesdropper}
\acro{csi}[CSI]{channel state information}
\acro{awgn}[AWGN]{additive white Gaussian noise}
\acro{iid}[i.i.d.]{independent and identically distributed}
\acro{ut}[UT]{user terminal}
\acro{bs}[BS]{base station}
\acro{mt}[MT]{mobile terminal}
\acro{eve}[Eve]{eavesdropper}
\acro{lse}[LSE]{least squared error}
\acro{mse}[MSE]{mean squared error}
\acro{glse}[GLSE]{generalized least squared error}
\acro{rls}[RLS]{regularized least-squares}
\acro{rhs}[r.h.s.]{right hand side}
\acro{lhs}[l.h.s.]{left hand side}
\acro{wrt}[w.r.t.]{with respect to}
\acro{tdd}[TDD]{time-division duplexing}
\acro{papr}[PAPR]{peak-to-average power ratio}
\acro{mrt}[MRT]{maximum ratio transmission}
\acro{zf}[ZF]{zero forcing}
\acro{rzf}[RZF]{regularized zero forcing}
\acro{srzf}[SRZF]{secure \ac{rzf}}
\acro{snr}[SNR]{signal to noise ratio}
\acro{sinr}[SINR]{signal to interference plus noise ratio}
\acro{rf}[RF]{radio frequency}
\acro{mf}[MF]{match filtering}
\acro{mmse}[MMSE]{minimum mean squared error}
\acro{rss}[RSS]{residual sum of squares}
\acro{amp}[AMP]{approximate message passing}
\acro{np}[NP]{non-deterministic polynomial time}
\acro{dca}[DCA]{DC programming algorithm}
\acro{lasso}[LASSO]{least absolute shrinkage and selection operator}
\end{acronym}

\end{document}